\documentclass[aps,twocolumn,showpacs,amsmath,amssymb]{revtex4} 
 
\usepackage{graphicx}
\usepackage{dcolumn}
\usepackage{bm}
\begin{document}

\title{Elliptic flow in nuclear collisions at the Large Hadron Collider} 
 
\author{ 
H. Niemi, K.J. Eskola and P.V. Ruuskanen} 
\email{harri.niemi, kari.eskola, vesa.ruuskanen@phys.jyu.fi} 
\affiliation{Department of Physics, 
P.O.Box 35, FIN-40014 University of Jyv\"askyl\"a, Finland\\ 
Helsinki Institute of Physics, 
P.O.Box 64, FIN-00014 University of Helsinki, Finland}

\date{June 6, 2008}

\begin{abstract} 
We use perfect-fluid hydrodynamical model to predict the elliptic flow
coefficients in Pb + Pb collisions at the Large Hadron Collider (LHC). The
initial state for the hydrodynamical calculation for central $A + A$ collisions
is obtained from the perturbative QCD + saturation (EKRT) model. The centrality
dependence of the initial state is modeled by the optical Glauber model. We show
that the baseline results obtained from the framework are in good agreement with
the data from the Relativistic Heavy Ion Collider (RHIC), and show predictions
for the $p_T$ spectra and elliptic flow of pions in Pb + Pb collisions at the
LHC. Also mass and multiplicity effects are discussed.
\end{abstract} 
 
\pacs{25.75.-q, 25.75.Dw, 25.75.Ld, 47.75.+f} 
 
\maketitle 

\section{Introduction}

Azimuthal anisotropies of hadron spectra are a good measure of collective
behavior of the dense particle system formed in the ultrarelativistic heavy ion
collisions \cite{Ollitrault:bk}. The origin of such anisotropies is thought to be
the rescattering among the particles, which are initially produced in isotropic
partonic interactions. These anisotropies can be quantified by the second
Fourier coefficient $v_2$, the so called elliptic flow coefficient, of the
azimuthal hadron distribution. In non-central Au + Au collisions at the
Relativistic Heavy Ion Collider (RHIC) at BNL the observed quite large $v_2$ is
regarded as one of the strongest signals pointing towards the formation of
thermalized strongly interacting partonic matter, the Quark-Gluon Plasma (QGP).

A common way to interpret the measured elliptic flow data is through full
equilibrium, perfect-fluid hydrodynamical models, see e.g. reviews
\cite{Kolb:2003dz, Huovinen:2003fa, Huovinen:2006jp}, with the Cooper-Frye
freeze-out mechanism \cite{Cooper:1974mv}. Consistency between these type of
models and the measured data at RHIC is regarded as evidence for low viscosity
\cite{Gyulassy:2004zy} and fast thermalization \cite{Kolb:2000sd} of the QGP.
However, a complete description of the low-$p_T$ spectra and elliptic flow
coefficients of pions simultaneously with e.g. those of protons is problematic
in this simple approach.

To improve the modeling, several different approaches have been developed. If
the chemical freeze-out is taken to happen before the complete kinetic
freeze-out in the Cooper-Frye type of decoupling \cite{Hirano:2002ds,
Kolb:2002ve}, a good simultaneous description of e.g. the transverse momentum
spectra of pions and protons can be obtained \cite{Huovinen:2007xh}. However,
elliptic flow tends to be overestimated in these models. At the moment, perhaps
the best agreement with the data is obtained from the hybrid models, which
combine hydrodynamical treatment of the QGP and hadron cascade simulation of the
hadronic interactions \cite{Teaney:2001av, Hirano:2007ei, Hirano:2005xf,
Nonaka:2006yn}. It is clear that a proper understanding of hadronic interactions
and the freeze-out mechanism are important in interpreting the observed data at
RHIC. Also hydrodynamical codes with shear viscosity corrections have been
recently developed for non-central collisions \cite{Romatschke:2007mq,
Song:2007ux, Luzum:2008cw, Dusling:2007gi, Chaudhuri:2008sj}. These calculations
indicate also low values of viscosity in the expanding matter.

The key input to the hydrodynamical models are the initial state and the
equation of state (EoS) of the QCD matter. When the initial state is obtained by
fitting the hydrodynamical model to the RHIC data, the observed total hadron
multiplicities fix the initial total entropy very well, but the transverse
profiles of the initial densities are not as well constrained, see e.g.
Ref.~\cite{Luzum:2008cw}. One possibility for controlling this uncertainty is to
use theoretically predicted initial conditions. In this approach it is possible
to predict the initial state for different collision energies and nuclei as
well. This has been our strategy e.g. in Refs.~\cite{Eskola:1999fc,
Eskola:2001bf, Eskola:2002wx, Eskola:2005ue}.

In \cite{Eskola:2005ue} we used the initial state from the EKRT final state
saturation model \cite{Eskola:1999fc} and showed that the perturbative QCD +
saturation + hydrodynamics approach gives a good description of the $p_T$
spectra of pions and kaons, and hadronic multiplicities in central Au + Au
collisions at RHIC. We also presented the predictions for the hadron $p_T$
spectra in central Pb + Pb collisions at the Large Hadron Collider (LHC). We
have also shown predictions for the elliptic flow at the LHC
\cite{Eskola:2007mx, Abreu:2007kv}. The aim of the present work is to expand our
previous studies of the hadron spectra and elliptic flow at the LHC and to probe
the uncertainties in the predictions. A similar approach with a calculated
initial state is used in Ref.~\cite{Hirano:2007gc}, where the initial state is
obtained from the Color-Glass Condensate model \cite{Iancu:2003xm}. Other LHC
predictions for the elliptic flow using the perfect-fluid hydrodynamics can be
found in Refs.~\cite{Abreu:2007kv, Bluhm:2007nu, Chojnacki:2007rq} and in the framework of
viscous hydrodynamics in Ref.~\cite{Chaudhuri:2008je}.

In this study we adopt the simple approach and use the perfect-fluid
hydrodynamics for the space-time evolution of the matter, assuming full kinetic
and chemical equilibrium throughout the evolution, see Sec.~II. In spite of its
restrictions discussed above, the full-equilibrium hydrodynamical approach, once
tested against the RHIC data, provides a good framework for baseline predictions
of the low-$p_T$ pion spectra and elliptic flow coefficients at the LHC. As
discussed in Sec.~III below, the EKRT model gives us the initial state for the
hydrodynamical evolution in central $A + A$ collisions for both RHIC and the LHC
energies. Extension to non-central collisions is made by using the optical
Glauber model at two different limits. The hydrodynamical evolution, elliptic
flow, eccentricities and transverse flow at RHIC and LHC are discussed in Sec.
IV. We will first, in Sec.~V, show that the model results agree with the RHIC
data, and then in Sec.~VI present our predictions for the transverse momentum
spectra of pions and for the elliptic flow coefficients of pions and protons at
the LHC. Conclusions are given in Sec.~VII.

\section{Hydrodynamical framework}

Once the initial energy density $\epsilon$ and the net-baryon density $n_B$ are
given and the Equation of State (EoS) $P=P(\epsilon,n_B)$ is known, the
evolution of matter can be described by relativistic hydrodynamics. By solving
the perfect-fluid hydrodynamic equations, i.e. the local conservation of
4-momentum, $\partial_{\mu}T^{\mu\nu}=0$, and of net-baryon number,
$\partial_{\mu} (n_Bu^{\mu})=0$, one obtains the space-time evolution of all
thermodynamic quantities and the collective flow velocity $u^{\mu}=\gamma(1,{\bf v_T},v_z)$ 
of the matter. When considering particle production at midrapidities,
where the rapidity spectra are approximately flat, these equations can be
simplified by assuming longitudinal boost invariance. In this case the
longitudinal flow velocity is given by $v_z = z/t$ \cite{Bjorken:1982qr}, and
all hydrodynamical quantities become independent of the space-time rapidity
$\eta = (1/2) \ln \left[\left(t+z\right)/\left(t-z\right)\right]$, i.e. they
depend on the transverse coordinates $x$ and $y$, and the longitudinal proper
time $\tau=\sqrt{t^2-z^2}$ only. This reduces the (3+1)-dimensional problem to a
(2+1)-dimensional one. We write then the conservation laws in these variables
and, including the EoS, solve them numerically by applying the SHASTA algorithm
\cite{Boris73, Zalesak79}.

In constructing the EoS \cite{Sollfrank:1996hd}, we describe the
high-temperature phase as an ideal gas of massless quarks and gluons with number
of flavors $N_f = 3$ and a bag constant $B$, while the low-temperature phase is
taken as an ideal gas of all hadronic states with $m < 2$ GeV \cite{PDG}. The
QGP and hadron resonance gas (HRG) phases are connected via the Maxwell
construction with mixed phase (MP) between the QGP and the HRG phase. The order
of the phase transition has been shown to have only small effects on the hadron
$p_T$ spectra and the elliptic flow of pions \cite{Huovinen:2005gy}. As in our
earlier studies \cite{Eskola:2002wx, Eskola:2005ue}, $B$ is chosen such that the
phase transition temperature is $T_c = 165$ MeV. We assume full kinetic and
chemical equilibrium for both phases throughout the temperature range considered
in this work.

Final hadron spectra are calculated through the Cooper-Frye decoupling procedure
\cite{Cooper:1974mv} as particle emission from a constant-$T$ surface obtained
from the hydrodynamic calculation. We determine the decoupling temperature
$T_{\rm dec}$ -- a parameter which controls the $p_T$ slopes of the hadron
spectra in the single-$T_{\rm dec}$ hydrodynamic framework  -- from the RHIC
data, as described in Refs. \cite{Eskola:2002wx, Eskola:2005ue}. For a recent
discussion of the relation of $T_{\rm dec}$ and decoupling dynamics, see
\cite{Eskola:2007zc, Heinz:2006ur}. After the Cooper-Frye decoupling, all 2- and
3-body strong and electromagnetic decays of unstable hadronic states are
accounted for. Thus, the feed-down from weak decays is not included in this
work.

\section{Initial state and centrality selection}

As initial conditions, the boost-invariant hydrodynamic calculation, at an
impact parameter $b$, requires the densities $\epsilon(x,y,\tau_0; b)$,
$n_B(x,y,\tau_0; b)$ at an initial time $\tau_0$. Our reference baseline is
central collisions, for which we obtain the initial densities from the EKRT
minijet (final state) saturation model \cite{Eskola:1999fc}. In addition to the
primary partonic transverse energy and the net-baryon number produced at
midrapidity in central AA collisions, the EKRT model also gives the average
formation time in terms of the saturation momentum, $\tau_f=1/p_{\rm sat}$. 

Assuming immediate thermalization at production, $\tau_0=\tau_f$, using the
binary collision (BC) profiles for the transverse-coordinate dependence of the
densities, and setting $T_{\rm dec}=150$~MeV, we have previously shown that the
minijet + saturation + hydrodynamics model is in good agreement with the RHIC
data in most central Au+Au collisions \cite{Eskola:2002wx, Eskola:2005ue}. We
have also demonstrated that using the wounded nucleon (WN) profiles and $T_{\rm
dec}=140$~MeV leads to practically equally good results \cite{Eskola:2007zc}.
Also predictions for the central Pb+Pb collisions at the LHC have been presented
\cite{Eskola:2005ue}.

The determination of the transverse profiles for $\epsilon$ and $n_B$ is,
however, problematic for the following reasons: First, the simplest version of
the EKRT model for head-on collisions  has only one saturation momentum scale,
$p_{\rm sat}$, which is perturbative, $\sim 1-2$~GeV for large nuclei at RHIC
and LHC. In the localized version \cite{Eskola:2000xq} $p_{\rm sat}$ depends on
the transverse location but near the edges of the system, where the produced
matter density becomes low enough, $p_{\rm sat}$ becomes non-perturbative and
the minijet calculation unreliable. Second, non-central collisions can be
expected to constitute a multiscale problem as very different nuclear regions
are colliding with each other, and we cannot expect the simple EKRT model to
describe the more peripheral collisions very well. Third, as discussed in
\cite{Eskola:2000xq, Eskola:2001rx}, different regions in the transverse plane
obviously form at different times; the center with a larger saturation scale
forms earlier than the edges of the system. This phenomenon adds the
complication of needing to provide the initial conditions on a surface
$\tau=\tau_0(x,y)$ instead of just at a fixed initial time $\tau=\tau_0$ (see
the discussion in \cite{Kolb:2001qz}). 

It is then obvious that some additional modeling is necessary in fixing the
initial transverse profiles of $\epsilon$ and $n_B$ for non-central collisions
in particular. One possibility is to assume a mixture of the BC and WN 
profiles, either for the energy density or the entropy density
\cite{Kolb:2001qz}, with the mixture coefficient determined from RHIC
data. Then, however, since the initial transverse profiles should be affected by
production dynamics as discussed in \cite{Eskola:2001rx}, an uncertainty in the
extrapolation to the LHC energies would remain anyway since the mixture
coefficients can change from RHIC to the LHC.

Instead of introducing further model details and thus also further model
uncertainties regarding the transverse profiles and their $\sqrt{s}$ dependence,
we choose a simpler and more transparent approach: We fix the initial time
$\tau_0$ to the value computed from the EKRT model in central collisions, and we
compute the spectra and elliptic flow coefficients with the BC and WN
initializations separately, considering these as two limiting cases of
transverse profiles both at RHIC and at the LHC. 

Our BC initial conditions for $A+A$ collisions are analogous to the eBC model of
Ref.~\cite{Kolb:2001qz} and are given by  
\begin{eqnarray}
\epsilon_{\rm BC}({\bf r}; {\bf b}) 
&=& \frac{1}{\tau_0}\frac{\left[dE/ d\eta\right]_{b=0}}{N^{AA}_{\rm BC}(0)} \frac {dN_{\rm BC}^{AA}} {d^2{\bf r} } ({\bf r};{\bf b})\cr
n_B^{\rm BC}({\bf r}; {\bf b}) &=& 
\frac{1}{\tau_0}\frac{\left[dN_B/ d\eta\right]_{b=0}}{N^{AA}_{\rm BC}(0)} \frac{ dN_{\rm BC}^{AA}({\bf b})} {d^2{\bf r}} ({\bf r};{\bf b}),
\label{eBC}
\end{eqnarray}
where $\left[dE/ d\eta\right]_{b=0}$ and $\left[dN_B/
d\eta\right]_{b=0}$ and $\tau_0$ are obtained from the EKRT model \cite{Eskola:2005ue} for
central collisions and where the BC profile is 
\begin{equation}
\frac{dN_{\rm BC}^{AA} } {d^2{\bf r} } ({\bf r};{\bf b})= T_A({\bf r}+\frac{\bf b}{2})T_A({\bf
r}-\frac{\bf b}{2}) \sigma_{NN}^{\rm in},
\end{equation} 
and $N^{AA}_{\rm BC}(0) = \sigma_{NN}^{\rm in}\int d^2\mathbf{r} \left[T_A({\bf r})\right]^2$. In
computing the standard nuclear thickness functions $T_A$, we use the Woods-Saxon
parametrization of the nuclear density. Similarly, our WN initial conditions are
analogous to the eWN model of Ref.~\cite{Kolb:2001qz} and computed from 
\begin{eqnarray}
\epsilon_{\rm WN}({\bf r}; {\bf b})&=& C_{\epsilon} \frac{dN_{\rm WN}^{AA}}{d^2{\bf r}}({\bf r};{\bf
b})\cr
n_B^{\rm WN}({\bf r}; {\bf b}) &=& C_B  \frac{dN_{\rm WN}^{AA}}{d^2{\bf r}}({\bf r};{\bf b}),
\label{eWN}
\end{eqnarray}
where the normalization constants $C_{\epsilon}$ and $C_B$ are fixed by
requiring, for central collisions, the initial entropy $dS/d\eta$ and the
initial net-baryon number in the eWN initial state to be the same as in the BC
initial state. The WN profile is given by 
\begin{eqnarray}
\frac{dN_{\rm WN}}{d^2{\bf r}}({\bf r};{\bf b}) \cr
= 
T_A({\bf r}+{\frac{\bf b}{2}}) &\big[1 - \bigg(1-\frac{\sigma_{NN}^{\rm in}}{A} T_A({\bf
r}-{ \frac{\bf b}{2}})\bigg)^A\big] \cr
+ 
T_A({\bf r}-{\frac{\bf b}{2}}) &\big[1 - \bigg(1-\frac{\sigma_{NN}^{\rm in}}{A}T_A({\bf
r}+{ \frac{\bf b}{2}})\bigg)^A\big].
\end{eqnarray}
The initial state parameters for the different collision energies are shown in
Table I. The integral of $dN_{WN}/d^2\mathbf{r}$ over the transverse plane is
the number of participants, $N_{\rm part}(\mathbf{b})$, for a collisions at
impact parameter $\mathbf{b}$.
\begin{table}[!h]
\centering
\begin{ruledtabular}
\begin{tabular}{lcc}	
								& RHIC & LHC	\\ 
$\sqrt{s_{NN}}$ [GeV] 						& 200 	& 5500	\\
$\tau_0$ [fm] 							& 0.17	& 0.097	\\
$\left[dE/d\eta\right]_{b=0}$ [GeV]				& 2460	& 14800	\\
$\left[dN_B/d\eta\right]_{b=0}$ 		 		& 15.3	& 3.36	\\
$\sigma^{in}_{NN}$ [mb] 		 			& 42	& 60	\\
\end{tabular}
\end{ruledtabular}
\caption{\protect\small  The initial state parameters for Au + Au collisions at RHIC
and Pb + Pb at the LHC. } 
\end{table}

In other words, we assume here that once the EKRT model fixes the normalizations
of $\epsilon$ and $n_B$ for central collisions, the optical Glauber model gives
the impact parameter dependence of the initial conditions, assuming
proportionality either to the BC or WN densities. We would like to emphasize
that in this work we consider the two limits as they are. We do not try to fit
the RHIC data by finding the best linear combination of the eBC and eWN
profiles. Instead, we show that the RHIC data fall between the two limits and
take the difference to represent the uncertainty in the extrapolation to the LHC
energy.

Qualitatively, there are two main differences between the two initializations.
First, the number of BCs drops much faster with increasing impact parameter than
the number of WNs. This leads to a faster dropping multiplicity for the eBC than
for the eWN initialization as a function of centrality. Second, at a given
impact parameter the BC density in the transverse plane falls faster as a
function of the transverse distance from the center of the fireball than the WN
density. At a given multiplicity, this leads to stronger transverse pressure
gradients for the eBC than for the eWN initialization. Thus, if the same
decoupling condition (same $T_{\rm dec}$) were used, the eBC initialization
would lead to both stronger transverse flow and larger elliptic flow than the
eWN initialization. 

To discuss the centrality classes, we apply the optical Glauber model, where
the total cross section is given by 
\begin{equation}
\sigma_{\rm tot}^{AA}= \int d^2{\bf b} \frac{d\sigma_{\rm tot}}{d^2{\bf b}} = 
	\int d^2{\bf b}[1-e^{T_{AA}({\bf b})\sigma_{NN}^{\rm in}}],
\end{equation}
where $T_{AA}$ is the standard nuclear overlap function. For the centrality
classes considered here, $c_1=0-5\%$, $c_2=5-10\%$, $c_3=10-15\%$, ...,  we find
the impact parameter ranges $[0,b_1]$, $[b_1,b_2]$, ... such that, e.g., for the
2nd centrality class $c_2$
\begin{equation}
\sigma_{c_2}^{AA}/\sigma_{\rm tot}^{AA} = 0.05
=\int_{b_{1}}^{b_2} d^2{\bf b} [1-e^{T_{AA}({\bf b})\sigma_{NN}^{\rm in}}],
\end{equation} 
and similarly for the other classes $c_i$. Using the $d\sigma_{\rm tot}/d^2b$ as
the weight, we determine the average impact parameter for each centrality class
as follows
\begin{equation}
\langle b \rangle_i = \frac{1}{\sigma_{\rm c_i}^{AA}} \int^{b_i}_{b_{i-1}} d^2{\bf b}  
[1-e^{T_{AA}({\bf b})\sigma_{NN}^{\rm in}}] b.
\end{equation} 
The average number of participants in a given centrality class is computed
similarly. The initial densities for each centrality class are then computed
using $b=\langle b \rangle_i$ in Eqs. \ref{eBC} and \ref{eWN}. The average
impact parameters and average numbers of participants for selected centrality
classes are shown in Table II.
\begin{table}[!h]
\centering
\begin{ruledtabular}
\begin{tabular}{cccccccc}	
	&&\multicolumn{2}{c}{RHIC}&&\multicolumn{2}{c}{LHC}\\ \cline{3-4} \cline{6-7}
centrality \%	&& $b$ [fm]	& $N_{\rm part}$ 	&& $b$ [fm] 	& $N_{\rm part}$	\\ \hline
0-5	&& 2.24		& 347	&	& 2.31	& 374	\\
5-10	&& 4.09		& 289	&	& 4.23	& 315	\\
10-15	&& 5.30		& 242	&	& 5.47	& 264	\\
15-20	&& 6.27		& 202  	&	& 6.48	& 221   \\
20-30	&& 7.49		& 153  	&	& 7.74	& 168   \\
30-40	&& 8.87		& 102  	&	& 9.17	& 112   \\
40-60	&& 10.6		& 50.8 	&	& 10.9	& 56.7  \\
60-70	&& 12.1		& 19.6 	&	& 12.5	& 21.2  \\
70-80	&& 13.0		& 9.13	& 	& 13.4	& 9.65  \\ 
\end{tabular}
\end{ruledtabular}
\caption{\protect\small  The average impact parameters and the average numbers of participants in selected centrality
classes from the optical Glauber model.} 
\end{table}

\section{Elliptic flow, eccentricities and transverse flow}

The transverse momentum and rapidity dependent Fourier coefficients $v_n$ for
each centrality class are defined as
\begin{equation}
v_n(y,p_T;b) \equiv 
\left(\frac{dN(b)}{dydp^2_T}\right)^{-1}
\int_{-\pi}^{\pi} d\phi\cos(n\phi)\frac{dN(b)}{dydp^2_Td\phi},
\end{equation}
while the $p_T$-integrated $v_n$'s are given by 
\begin{equation}
v_n(y;b) \equiv 
\left(\frac{dN(b)}{dy}\right)^{-1}
\int_{-\pi}^{\pi} d\phi\cos(n\phi)\frac{dN(b)}{dyd\phi},
\end{equation}
where $b$ stands for the average impact parameter $\langle b \rangle_i$ in the
centrality class $c_i$. Due to the longitudinal boost symmetry assumed here, the
$v_n$ coefficients, which we compute from the hydrodynamic spectra, do not
depend on rapidity, thus $y=0$ is implicit. Our definition of the minimum bias
elliptic flow coefficient, $v_2^{\rm m.bias}(y, p_T)$, in turn is as follows,
\begin{equation}
v_2^{\rm m.bias}(y, p_T) \equiv
\frac
{\int d^2{\rm b}v_2(y, p_T; b)\frac{dN(b)}{dydp^2_T}}
{\int d^2{\rm b}\frac{dN(b)}{dydp^2_T}}.
\end{equation}
In the calculations the impact parameter is in the $x$ direction. Therefore, the
system is initially shorter and the pressure gradients stronger in the $x$
direction. Stronger pressure gradients will generate stronger transverse flow
during the evolution in the short direction. As a result the system grows faster
in the $x$ direction than in the $y$ direction. The generated flow asymmetry
manifests in the azimuthal dependence of hadron spectra and a non-zero $v_2$ is
observed.

Figures 1 and 2 show the QGP-MP and MP-HRG phase boundaries and the decoupling
boundaries in the $x$ and $y$ directions from our hydrodynamical simulations
with $b = 7.49$ fm for RHIC and $b = 7.74$ fm for the LHC, both corresponding to
the $20-30$\% centrality class. Both the RHIC and LHC results are shown with two
different initializations, the eBC and eWN, discussed in the previous section.
Lifetimes of the different phases at different transverse locations can be read
off from Figs.~1 and 2 in each case. With the same collision
energy different initial profiles correspond to slightly different multiplicity.
Hence there is a difference e.g. in the initial transverse size of the eBC and
eWN initializations in the figures. The first-order phase transition, present in
the EoS we use, produces a shock wave at the boundary of HRG and MP, which in
turn causes the peaks at the edge of the system seen especially in Fig.
2. However, contribution to the hadronic observables from the peaks is small. We
have checked that removing the peaks causes less than 5\% changes in the hadron
spectra and elliptic flow coefficients.

\begin{figure}[bht] 
\vspace{-0.0cm} 
\hspace{-0.0cm} 
\includegraphics[width=9.3cm]{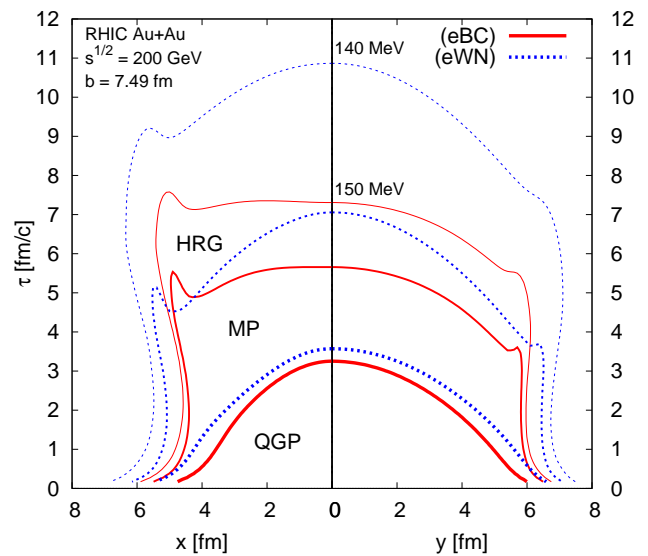} 
\vspace{-0.2cm} 
\caption{\protect\small (Color online) The phase boundaries and the constant-$T_{\rm dec}$ 
decoupling curves from our calculation for $\sqrt{s_{NN}}=200$~GeV  Au + Au collisions 
at RHIC in the 20-30 \% centrality class. Both $x$ and $y$ directions are shown~\cite{Song:2007ux}. }
\end{figure} 
\begin{figure}[bht] 
\vspace{-0.0cm} 
\hspace{-0.0cm} 
\includegraphics[width=9.0cm]{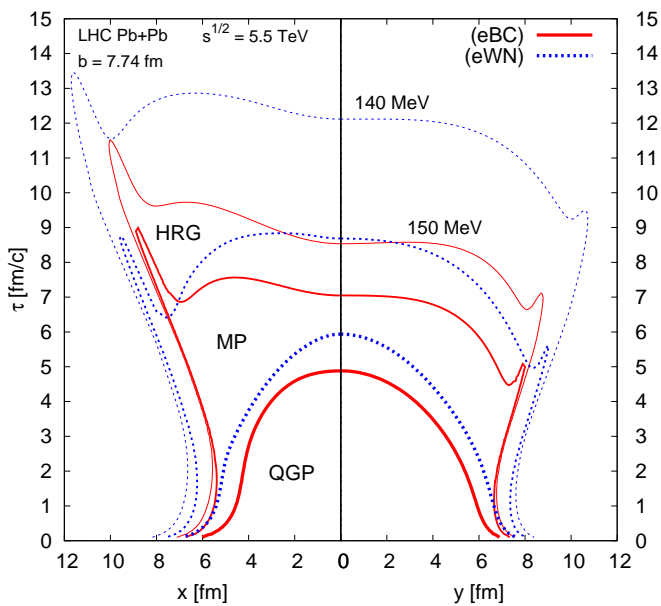} 
\vspace{-0.2cm} 
\caption{\protect\small (Color online) The Phase boundaries and the constant-$T_{\rm dec}$ MeV 
decoupling curves for $\sqrt{s_{NN}}=5500$~GeV Pb + Pb collisions at the LHC
in the 20-30 \% centrality class.} 
\end{figure} 

A convenient way to illustrate the global features of the hydrodynamical
space-time evolution of the matter is through the following three quantities:
spatial eccentricity, momentum-space eccentricity and average transverse flow.
These quantities are introduced in Ref.~\cite{Kolb:2000sd}, where also their
behavior for different systems has been discussed. The spatial eccentricity is
defined as
\begin{equation}
\varepsilon_{x} \equiv \frac{\langle y^2 - x^2 \rangle}{\langle y^2 + x^2 \rangle} \equiv \frac{\int
dxdy\,\epsilon(x, y, \tau) (y^2 - x^2)}{\int dxdy\,\epsilon(x, y, \tau) (y^2 + x^2)},
\end{equation}
where the integral is over the transverse plane and the energy density $\epsilon$ is used as the
weighting factor. Similarly, the momentum space eccentricity is defined as
\begin{equation}
\varepsilon_{p} \equiv \frac{\int dxdy\, (T^{xx} - T^{yy})}{\int dxdy\,(T^{xx} + T^{yy})},
\end{equation}
where $T^{ii}$ are the components of the energy-momentum tensor. The average transverse flow can be
defined as
\begin{equation}
\langle v_T \rangle \equiv \frac{\langle \gamma v_T \rangle}{\langle \gamma \rangle} \equiv
\frac{\int dxdy\,\epsilon(x, y, \tau) \gamma v_T(x, y, t)}{\int dxdy\,\epsilon(x, y, \tau) \gamma},
\end{equation}
where $v_T = \sqrt{v_x^2+v_y^2}$ and the $\gamma$ factor has been used as an additional weight. 

The spatial eccentricity $\varepsilon_x$ measures the asymmetry of the matter
distribution in the $x$ and $y$ directions, thus it also measures the asymmetry
of pressure gradients. In a noncentral heavy ion collision, with the impact
parameter along the $x$-axis, the initial distribution of the matter corresponds
to a positive initial $\varepsilon_x$ which acts as a driving force to an
asymmetric transverse flow. The generated flow asymmetry is quantified by
$\varepsilon_p$. The sign convention of $\varepsilon_x$ and $\varepsilon_p$ is
such that an initially positive $\varepsilon_x$ leads to a positive
$\varepsilon_p$. A positive $\varepsilon_p$ at the end of the evolution will
convert to a positive elliptic flow coefficient $v_2$, which can be extracted
from hadron spectra. The value of $\varepsilon_p$, when the system decouples, is
approximately twice the $p_T$-integrated $v_2$ for pions \cite{Kolb:2000sd}, see
Figs. 3 and 5. During the evolution the established asymmetric flow field tends
to drive $\varepsilon_x$ towards zero, i.e. towards azimuthally symmetric matter
distribution. Therefore the driving force for the growth of elliptic flow is
strongest during the early stages of the collision. At later times
$\varepsilon_x$ goes through zero and eventually to slightly negative values and
$\varepsilon_p$ will not increase anymore.

\begin{figure}[bht] 
\vspace{-0.0cm} 
\hspace{-0.0cm} 
\includegraphics[width=8.5cm]{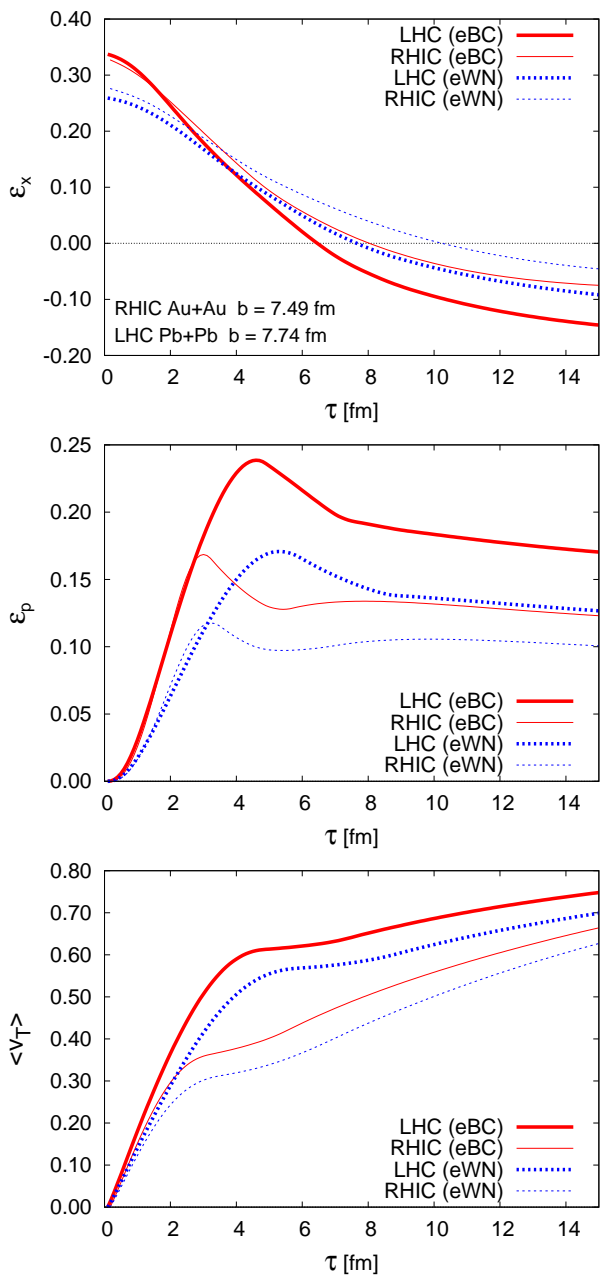} 
\vspace{-0.2cm} 
\caption{\protect\small (Color online) The spatial and momentum space eccentricity and the average transverse velocity 
as function of the longitudinal proper time for $\sqrt{s_{NN}}=200$~GeV  Au+Au collisions at RHIC (thin solid
and dashed curves) and $\sqrt{s_{NN}}=5500$~GeV Pb+Pb collisions at the LHC (thick solid and dashed curves)
in the 20-30 \% centrality class.} 
\end{figure} 

Figure 3 shows the time evolution of $\varepsilon_x$, $\varepsilon_p$ and
$\langle v_T \rangle$ from the same calculation as the curves in Figs. 1 and 2.
Initially $\varepsilon_x$ is very similar for RHIC and the LHC if the same
initialization profiles (eWN or eBC) are used. At the beginning of the
evolution, when most of the matter is in the QGP phase, the flow generated by
the pressure gradients starts to decrease $\varepsilon_x$. Simultaneously
$\varepsilon_p$ is increasing. The initial rates of change for $\varepsilon_x$
and $\varepsilon_p$ are very similar at RHIC and the LHC. During the QGP phase
there is no significant difference in the behavior of $\varepsilon_x$ and
$\varepsilon_p$ between the two collision energies, but much bigger difference
between the two chosen initial profiles. When the system enters the mixed phase
all pressure gradients vanish ceasing the increase of $\varepsilon_p$ and
$\langle v_T \rangle$, i.e. $\varepsilon_p$ saturates even before
$\varepsilon_x$ goes to zero. During the mixed phase the matter is not
accelerating, but the established flow field still expands the system
transversally. This will cause $\varepsilon_p$ actually to decrease during the
mixed phase. The transition times between the different phases for different
initializations can be read off from Figs. 1 and 2. These times coincide with
the structures seen in the behavior of $\varepsilon_p$, and $\langle v_T
\rangle$. Transitions between the phases happen at different times at different
transverse locations, therefore e.g. $\langle v_T \rangle$ never completely
saturates when most of the matter is in the mixed phase: part of the matter is
always either in the HRG or in QGP phase, where the pressure gradients do not
vanish.

A significant difference between the LHC and RHIC is the lifetime of the QGP
phase. At the LHC $\varepsilon_p$ has more time to grow before the system enters
the mixed phase, which leads to a larger $\varepsilon_p$ at the end of the
evolution. Therefore we also expect that the elliptic flow coefficient $v_2$
will be larger at the LHC than at RHIC. Longer lifetime of the QGP phase
reflects also in the behavior of $\varepsilon_x$ and $\langle v_T \rangle$. At
RHIC $\varepsilon_p$ saturates well before $\varepsilon_x$ goes to zero, and
$\varepsilon_x$ is still positive when the system enters the HRG phase and
$\varepsilon_p$ is slightly increasing. Therefore there could still be some
elliptic flow generated during the HRG phase at RHIC, i.e. $v_2$ would have some
sensitivity on the decoupling condition. This can be verified by an explicit
calculation and within our framework $v_2$ increases by $\sim 20-30$\% at
$N_{\rm part} \sim 150$ when the decoupling temperature is changed from $T_{\rm
dec} = 160$ MeV to $T_{\rm dec} = 130$ MeV. On the other hand at the LHC the
saturation of $\varepsilon_p$ and the sign change of $\varepsilon_x$ happen more
or less simultaneously and when the system enters the HRG phase $\varepsilon_x$
is already negative and $\varepsilon_p$ is slightly decreasing. At the LHC $v_2$
is slightly decreasing when $T_{\rm dec}$ is decreasing but the HRG effects are
significantly smaller at the LHC than at RHIC. Thus the predictions for the
$p_T$-integrated $v_2$ for the LHC are insensitive to the details of the HRG
dynamics and therefore quite robust once the initial profile is fixed.

In contrast to the behavior of $\varepsilon_p$, $\langle v_T \rangle$ never
completely saturates. Even at late stages of the collision $\langle v_T \rangle$
can still increase significantly. Although the azimuthal asymmetry of the pressure
gradients almost vanishes, the pressure gradient in the radial direction does not
vanish. Even though the gradient is small in the HRG phase, additional
transverse velocity is generated and can be easily seen in the slopes of hadron $p_T$
spectra. Thus the $p_T$ spectra of hadrons remain sensitive to the decoupling
temperature and HRG dynamics both at RHIC and the LHC. During the QGP phase the
pressure gradients are strongest and the longer lifetime of the QGP phase at the
LHC leads to a clearly larger transverse flow before the mixed phase than at
RHIC. Regardless of the decoupling condition the $p_T$ spectra at the LHC are
always flatter than at RHIC.

The behavior of the $p_T$-integrated $v_2$ for different initializations can be
quite easily determined from the behavior of $\varepsilon_p$ alone.
The differential elliptic flow coefficient $v_2(p_T)$ however depends not only on
$\varepsilon_p$, but also on the slopes of the hadron $p_T$ spectra, which in
turn depend on $\langle v_T \rangle$ and temperature. In general, increasing
$\varepsilon_p$ will increase $v_2(p_T)$ at fixed $p_T$ but increasing $\langle
v_T \rangle$ will decrease it \cite{Huovinen:2001cy}. The net effect depends on
the details of the HRG dynamics. Within our framework with full chemical and
kinetic equilibrium in the HRG throughout the evolution, the net result is that
$v_2(p_T)$ is quite insensitive to the decoupling temperature. At RHIC
$v_2(p_T)$ is slightly increasing and at the LHC slightly decreasing with
decreasing $T_{\rm dec}$. These changes are, however, small and practically
within our framework,  predictions for $v_2(p_T)$ are independent of the
decoupling temperature. Changes in the HRG dynamics would change this behavior,
e.g. chemical freeze-out before kinetic freeze-out \cite{Hirano:2002ds,
Kolb:2002ve, Huovinen:2007xh} would modify the dependence of spectral slopes on
temperature and flow conditions and thus would affect the behavior of
$v_2(p_T)$ as a function of the decoupling condition. These effects are,
however, not studied in this work.

\section{Results for RHIC }
\begin{figure}[bht] 
\vspace{-0.0cm} 
\hspace{-0.0cm} 
\includegraphics[height=9.0cm]{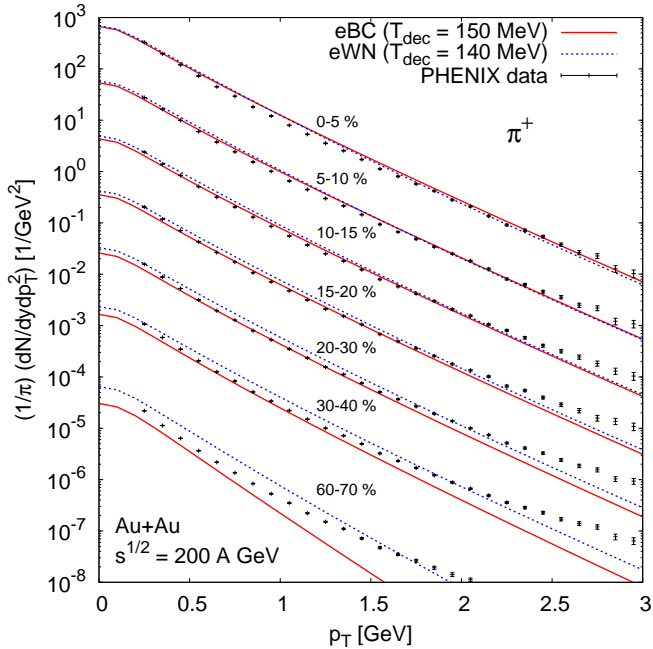} 
\vspace{-0.8cm} 
\caption{\protect\small (Color online) The $p_T$ spectra of positive pions for
$\sqrt{s_{NN}}=200$~GeV Au + Au collisions at RHIC compared with the PHENIX data 
\cite{Adler:2003cb}. The solid (dashed) lines are for the eBC (eWN) initialization.
The centrality classes are indicated in the figure and are scaled by the increasing
powers of $10^{-1}$.}
\end{figure} 
Figure 4 shows the calculated pion spectra for different centrality classes in
$\sqrt{s_{NN}} = 200$ GeV Au+Au collisions at RHIC, compared with the PHENIX
data \cite{Adler:2003cb}. Shown in the figure are the results corresponding to
the two different initializations we consider here. The decoupling temperature
is fixed to $T_{\rm dec}=150$~MeV for the eBC initialization, as in our previous
works \cite{Eskola:2002wx, Eskola:2005ue}. As seen in Fig.~3, the eWN
initialization generates less transverse flow for a given decoupling condition,
therefore -- in order to reproduce the pion $p_T$ spectra --  the system must be
allowed to decouple later than for the eBC initialization. In the eWN case,
$T_{\rm dec} = 140$~MeV describes the data well. With this difference in $T_{\rm
dec}$, both initializations give an equally good agreement with the data for
central and mid-central collisions. For more peripheral collisions the
calculations start to separate and, as expected, the eBC results fall below the
eWN results but the data lies still well between the two limits considered.
Fig.~4 also demonstrates how in most central collisions the hydrodynamically
computed pion spectra reproduce the data well in the region $p_T\lesssim 3$~GeV,
while the applicability region of hydrodynamic spectra in the 60-70 \%
centrality class is limited to $p_T\lesssim 1.5$~GeV, only. 

\begin{figure}[bht] 
\vspace{-0.0cm} 
\hspace{-0.0cm} 
\includegraphics[height=9.0cm]{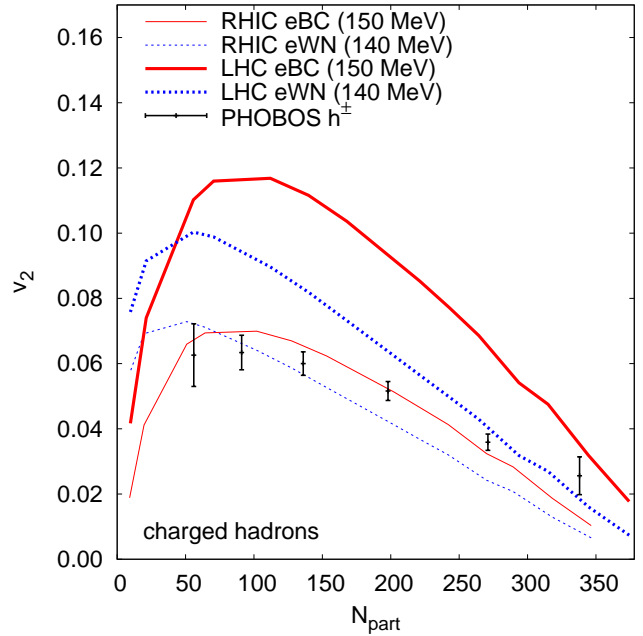} 
\vspace{-0.8cm} 
\caption{\protect\small (Color online) The $p_T$-integrated $v_{2}$ of charged hadrons for $\sqrt{s_{NN}}=200$~GeV 
Au + Au collisions at RHIC (thin lines) and $\sqrt{s_{NN}}=5500$~GeV Pb + Pb collisions at the LHC (thick lines)
vs. the number of participants. The data is from the PHOBOS collaboration \cite{Alver:2006wh}.} 
\end{figure} 
Figure 5 shows our model calculations of the $p_T$-integrated $v_2$ for charged
hadrons as a function of the number of participants, compared with the PHOBOS
data \cite{Alver:2006wh}. For mid-peripheral collisions ($N_{\rm part}\sim
150-200$) the agreement with the data is quite good. However, for the most central
classes we underestimate the data --  a typical feature also in
several other hydrodynamical models. The underestimation of $v_2$ is usually
associated with fluctuations in the initial geometry of the system
\cite{Miller:2003kd}, which are not accounted for in our treatment. For very
peripheral collisions the model eventually starts to overshoot the data but 
in these collisions, with the small system sizes and particle numbers, we 
expect, in any case, the validity of hydrodynamic modeling to deteriorate. 
From mid-peripheral to central collisions the eBC initialization
generates more elliptic flow than the eWN initialization, as expected on the
basis of Fig.~3. This results from the larger pressure gradient in the eBC
initialization. However, for more peripheral collisions the behavior changes
and the eBC results fall below the eWN ones. Since the total multiplicity for
the eBC initial state drops faster with increasing impact parameter, it also
causes the lifetime of the QGP phase to drop faster. For very peripheral
collisions the lifetime of the eBC system is considerably shorter than that of
the eWN system. Thus the elliptic asymmetry will not have time to develop before
freeze-out. Also the pressure gradients in the eBC case are larger than in the eWN
case for central and mid-peripheral collisions, but become very similar for
peripheral collisions. As a result, the elliptic flow predicted by the eBC
initialization will eventually drop below the eWN results as we go towards more
peripheral collisions. At RHIC the results coincide when $N_{\rm part}\sim
70$. In addition to the RHIC results, Fig.~5 shows also the LHC prediction,
which we shall comment on shortly.
\begin{figure}[bht] 
\vspace{-0.0cm} 
\hspace{-0.0cm} 
\includegraphics[height=9.0cm]{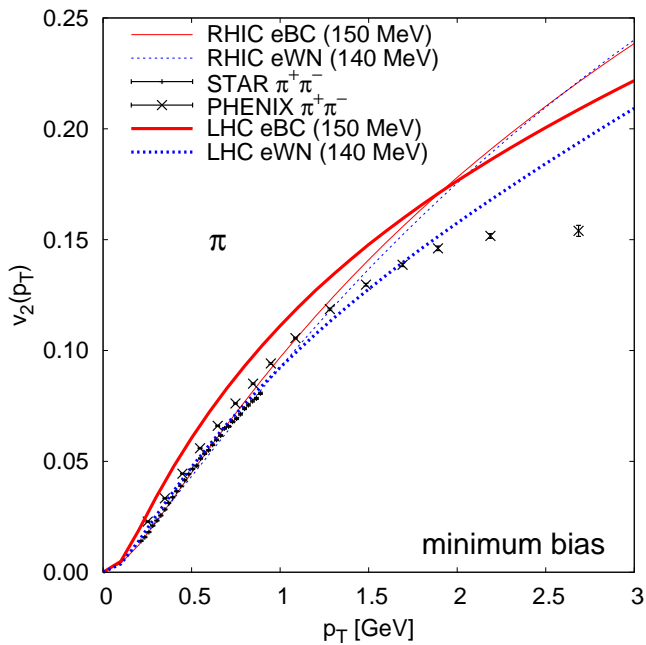} 
\vspace{-0.8cm} 
\caption{\protect\small (Color online) The minimum bias $v_{2}(p_T)$ of pions for $\sqrt{s_{NN}}=200$~GeV 
Au + Au collisions at RHIC (thin lines) and $\sqrt{s_{NN}}=5500$~GeV Pb + Pb collisions 
at the LHC (thick lines) as a function of transverse momentum. The data is from the PHENIX 
\cite{Adare:2006ti} and STAR \cite{Adams:2004bi} collaborations.} 
\end{figure} 

Figure 6 shows our results for the minimum bias $v_{2}(p_T)$ of pions in Au+Au collisions
at RHIC (thin lines), compared with the RHIC data from STAR \cite{Adams:2004bi}
and PHENIX \cite{Adare:2006ti}. The agreement with the data is quite good up to
$p_T \lesssim 1.5$ GeV for both initial profiles. As is the case with the
$p_T$-integrated $v_2$, also $v_2(p_T)$ is larger for the eBC than the eWN
initialization from mid-peripheral to central collisions. However, for
peripheral collisions also $v_2(p_T)$ with the eBC initialization eventually
falls below the eWN results for the reasons discussed above. This can be seen
in Fig.~5 as the crossing of the eWN and eBC curves for RHIC. For the minimum bias
$v_2(p_T)$ all centrality classes between $0-80$\% are included and the net
result is that $v_2(p_T)$ averages nearly to the same values with both
initializations at RHIC.

\begin{figure}[bht] 
\vspace{-0.0cm} 
\hspace{-0.0cm} 
\includegraphics[height=8.9cm]{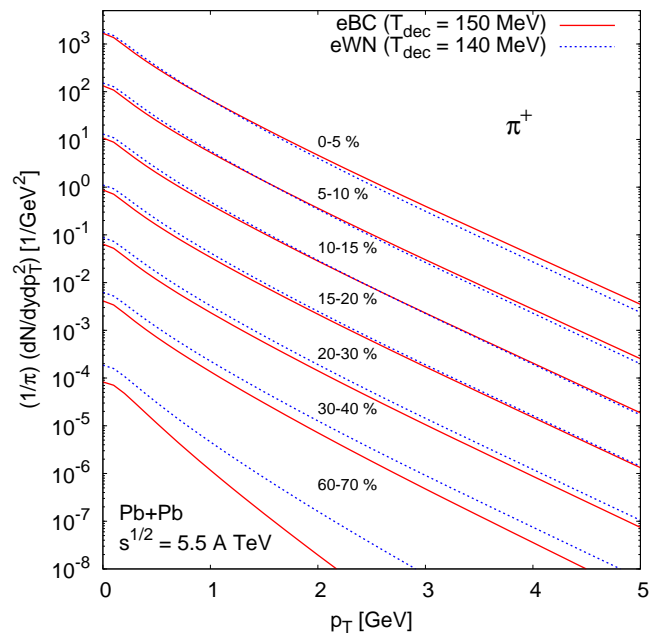} 
\vspace{-0.8cm} 
\caption{\protect\small (Color online) Our predictions for the transverse momentum spectra of positive pions for $\sqrt{s_{NN}}=5500$~GeV 
Pb + Pb collisions at the LHC. The results with the eBC (eWN) initialization are shown by solid
(dashed) lines. The centrality classes are the same as for RHIC in Fig.~4.} 
\end{figure} 

\section{Predictions for the LHC}
In Fig.~7 we show our prediction for the transverse momentum spectra of positive
pions for different centrality classes in $\sqrt{s_{NN}} = 5.5$ TeV Pb+Pb
collisions at the LHC. The charged hadron multiplicity in the most central
collisions is ca. $2600$~\cite{Eskola:2005ue}. Once the framework is tested in
RHIC Au + Au collisions, the only freedom left is the choice of the initial
profile and of the decoupling temperature $T_{\rm dec}$. For the initial profile
we consider the same limits as at RHIC, i.e. we do calculations with both the
eWN and eBC profiles, with normalization fixed from the EKRT model for central
collisions. As discussed in the context of dynamical decoupling
\cite{Eskola:2007zc}, we do not expect $T_{\rm dec}$ to change significantly
from RHIC to the LHC. Thus we show the eBC results with $T_{\rm dec} = 150$~MeV
and the eWN results with $T_{\rm dec} = 140$~MeV. Both initializations give
similar results for central and mid-peripheral collisions but, similarly to the
RHIC case, the two calculations start to separate for more peripheral
collisions.

In addition to the RHIC results, we show our predictions for the LHC in Fig.~5
for the $p_T$-integrated $v_2$ and in Fig.~6 for the minimum bias
$v_2(p_T)$, both for the eBC and eWN initializations. From Fig.~5 we see that,
as expected, the $p_T$-integrated $v_2$ is much larger at the LHC than at RHIC.
Due to the longer lifetime of the QGP at the LHC the flow asymmetry has more
time to develop before the system goes into the mixed phase where pressure
gradients vanish.

Although the $p_T$-integrated $v_2$ is clearly larger at the LHC than at RHIC,
the situation is not as clear for the minimum bias $v_2(p_T)$. For the eBC
initialization our LHC prediction is clearly above all the RHIC data points and,
in the small-$p_T$ region, also above the RHIC predictions. On the other hand,
in the low-$p_T$ region the eWN results for the LHC are very close to the
$v_2(p_T)$ values at RHIC. This does not contradict the results for the
$p_T$-integrated $v_2$, which are obtained by weighting with the particle's
$p_T$ spectra. Since the spectra at the LHC are much flatter than at RHIC (,i.e.
the average $p_T$ is larger at the LHC), $v_2$ gets more weight from the high
$p_T$ region at the LHC. Thus, even if the $v_2(p_T)$ curves coincide, the
$p_T$-integrated $v_2$ will be larger at the LHC. The difference in $v_2(p_T)$
between the RHIC and the LHC results is that at the LHC the differential $v_2$
is more sensitive to the initial profile than at RHIC, i.e. the eBC
initialization leads to clearly larger values of $v_2(p_T)$ than the eWN
initialization, while at RHIC both initializations give very similar results.
The reason for this is the larger overall multiplicity and thus longer lifetime
of the QGP phase at the LHC. Even in more peripheral collisions the lifetime of
the QGP phase is still long enough to generate enough elliptic flow such that
the ordering of $v_2(p_T)$ between the eBC and eWN initialization remains down
to $N_{\rm part} \sim 50$. This can be seen in Fig.~5.
\begin{figure}[bht] 
\vspace{-0.0cm} 
\hspace{-0.0cm} 
\includegraphics[height=9.0cm]{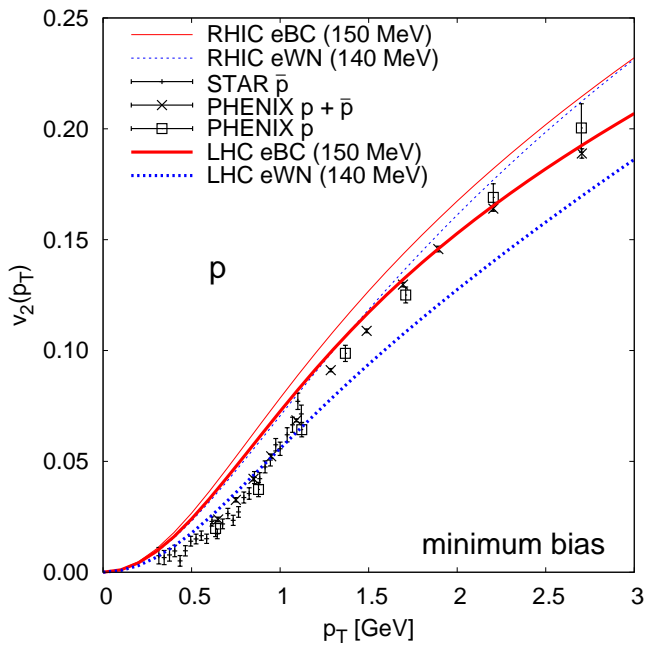} 
\vspace{-0.8cm} 
\caption{\protect\small (Color online) The minimum bias $v_{2}(p_T)$ of protons for $\sqrt{s_{NN}}=200$~GeV 
Au + Au collisions at RHIC (thin lines) and $\sqrt{s_{NN}}=5500$~GeV 
Pb + Pb collisions at the LHC (thick lines) vs. $p_T$. The RHIC data are from the STAR \cite{Adams:2004bi} 
and PHENIX \cite{Adare:2006ti, Adler:2003kt} 
collaborations.} 
\end{figure} 

Although our simple single-$T_{\rm dec}$ freeze-out model cannot reproduce in
detail the measured proton $p_T$-spectra and $v_2$ at RHIC, it is still
interesting to compare the general features in the calculated results for RHIC
and the LHC. Figure 8 shows the results from the same calculation as for Fig.~6,
but now for protons. Also the RHIC data from STAR \cite{Adams:2004bi} and PHENIX
\cite{Adare:2006ti, Adler:2003kt} collaborations are shown. When the
calculations are compared with each other, for both the eBC and eWN initial
profiles it is seen that the proton $v_2(p_T)$ at fixed $p_T$ at the LHC is
always \emph{below} the RHIC values. This can be understood as a mass effect \cite{Huovinen:2001cy}.
For the pions, the increase in average $p_T$ compensates the increase in the
flow asymmetry. However, for protons the increase of average $p_T$ is larger due
to the larger mass of the proton. Thus the increase of $\langle p_T \rangle$
overcompensates the increase in the flow asymmetry. Although we cannot make a
quantitative prediction of $v_2(p_T)$ for the protons, we can infer from the
results and arguments above that $v_2(p_T)$ at the LHC is expected to be below
the values measured at RHIC, while,  as in the case of pions, the
$p_T$-integrated $v_2$ is still expected to be larger at the LHC than at RHIC.

Finally, we study the uncertainty from multiplicity in predicting the behavior
of elliptic flow coefficients from RHIC to the LHC. Motivated by the
multiplicity predictions in Refs.~\cite{Abreu:2007kv, Armesto:2008fj}, we have
repeated the calculation for the LHC with half the multiplicity predicted by the
EKRT model, by adjusting 
\begin{eqnarray}
\tau_0 &\rightarrow& \sqrt{2} \tau_0 \\
\left[dE/d\eta\right]_{b=0} &\rightarrow& \frac{1}{2\sqrt{2}}\left[dE/d\eta\right]_{b=0},
\end{eqnarray}
as suggested by the saturation conjecture \cite{Eskola:1999fc}. Figure 9 shows
the integrated $v_2$ for the two different multiplicities at the LHC together
with results for RHIC and Fig.~10 the minimum bias $v_2(p_T)$. With the lower
multiplicity the lifetime of the QGP phase decreases and less elliptic flow is
generated, which is clearly seen in Fig.~9, where the integrated $v_2$ drops
halfway between the RHIC results and the original LHC prediction. The drop is
similar between the eBC and eWN initializations. If a low multiplicity is
observed at the LHC, the prediction with the eWN profile for $v_2$ extends down
to similar values as observed at RHIC. The integrated $v_2$ is clearly sensitive
to the observed multiplicity. The minimum bias $v_2(p_T)$, shown in Fig.~10 is
not as sensitive to the multiplicity as the integrated elliptic flow, but there
is still a visible difference as compared with the high-multiplicity prediction.
Already for the high multiplicity calculation with the eWN profile, the LHC
prediction is very close to the RHIC data and our calculations for RHIC. When
the LHC multiplicity is lowered the eWN result drops  at low $p_T$ even slightly
below the RHIC calculations and data. Also the eBC prediction drops at low $p_T$
closer to the RHIC results.
\begin{figure}[bht] 
\vspace{-0.0cm} 
\hspace{-0.0cm} 
\includegraphics[height=9.0cm]{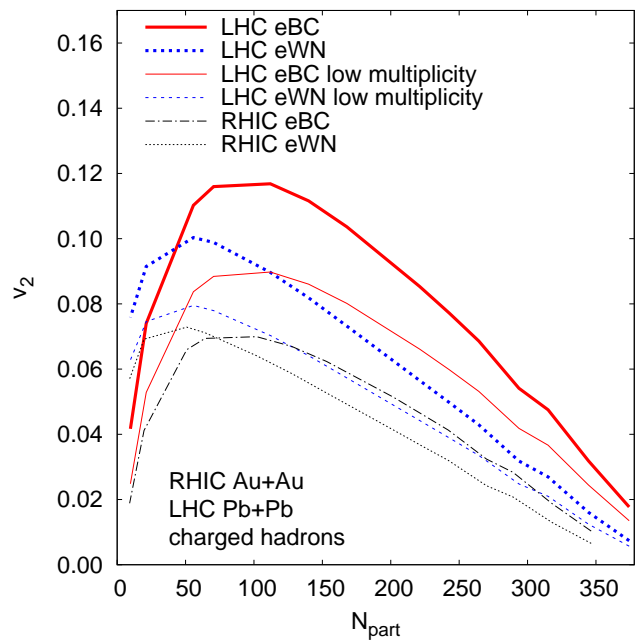} 
\vspace{-0.8cm} 
\caption{\protect\small (Color online) The integrated $v_2$ for the LHC low-multiplicity calculation} 
\end{figure} 
\begin{figure}[bht] 
\vspace{-0.0cm} 
\hspace{-0.0cm} 
\includegraphics[height=9.0cm]{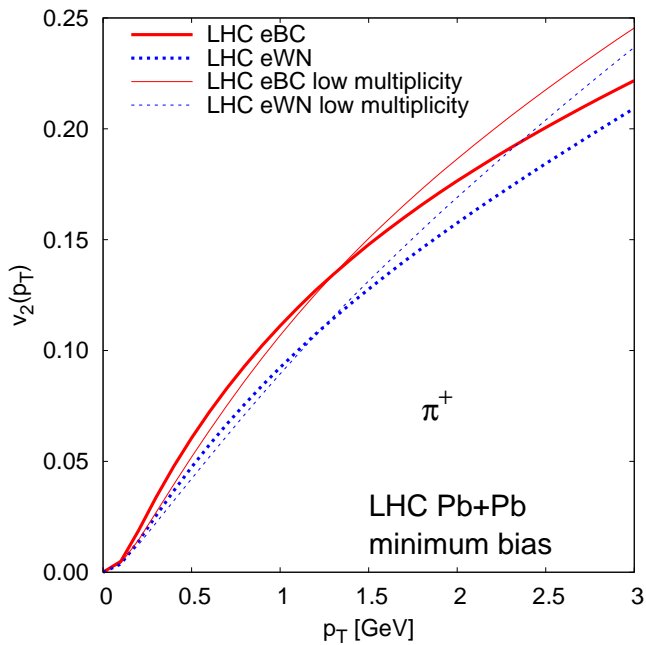} 
\vspace{-0.8cm} 
\caption{\protect\small (Color online) The minimum bias $v_2(p_T)$ for the LHC low-multiplicity calculation} 
\end{figure} 

\section{Conclusions}

We study elliptic flow in ultrarelativistic heavy ion collisions at RHIC and LHC
energies using the framework of perfect-fluid hydrodynamics. The initial state
for central heavy ion collisions is calculated from the pQCD + saturation model
\cite{Eskola:1999fc} and perfect-fluid hydrodynamics is used to model the
space-time evolution of the initially formed matter. Centrality dependence is
modeled by the optical Glauber model. We have shown that the results from this
approach are in reasonably good agreement with the measured data at RHIC, and
thus provide a good basis to predict both the pion spectra and elliptic flow
coefficients in $\sqrt{s_{NN}} = 5.5$ TeV Pb + Pb collisions at the LHC. The
main uncertainty in our model is the initial transverse profile of the energy
density. This uncertainty is addressed by considering two limits of the Glauber
model, taking the energy density proportional either to the density of binary
collisions or wounded nucleons. Three main observations are made for the LHC.
First we note that the $p_T$-integrated $v_2$ is expected to be much larger at
the LHC than at RHIC, since the lifetime of the QGP phase will be longer and
thus the pressure gradients that generate elliptic flow will persist longer. The
second observation is that $v_2(p_T)$ for pions will be unchanged or may
increase slightly from the values measured at RHIC. This can be understood as an
interplay between the flow asymmetry and flow magnitude \cite{Huovinen:2001cy}.
As the third point we observe that for heavier particles like protons,
$v_2(p_T)$ will be below the values measured at RHIC, even if the
$p_T$-integrated $v_2$ is larger. This is due to the fact that the increased
flow velocity is more important for protons, and thus it overcompensates the
increased flow asymmetry. 

We further noted that the $p_T$ integrated $v_2$ has still some sensitivity to the
decoupling condition and therefore to the HRG dynamics at RHIC, i.e. that
spatial eccentricity is not entirely converted to the momentum space
eccentricity before the matter enters the HRG phase. On the contrary, the
prediction of $v_2$ at the LHC is more robust than at RHIC, once the initial
profile and multiplicity are fixed: we do not expect the HRG dynamics to affect
significantly the integrated $v_2$ at the LHC. The differential elliptic flow
coefficient $v_2(p_T)$ was in turn found to be insensitive to the decoupling
condition at both collision energies. This insensitivity presumably follows from
our full equilibrium treatment of the HRG phase. Different HRG dynamics would
change this behavior. However, when the decoupling condition is fixed at RHIC
to reproduce the hadron spectra, the freeze-out temperature is not expected to
change significantly from RHIC to the LHC \cite{Eskola:2007zc}. Once the EoS and
the freeze-out parameters are fixed at RHIC, we expect that the same parameters
apply also at the LHC. Once the multiplicities are fixed, the
initial transverse profile remains as the largest uncertainty in the model.

The effect of multiplicity on the elliptic flow coefficients at the LHC is
studied by considering half the multiplicity that is predicted by the EKRT
model. It is found that the integrated $v_2$ is quite sensitive to the
multiplicity. The lower limit at the LHC, predicted by the eWN initialization,
comes even slightly below the upper limit at RHIC given by the eBC
initialization. The differential $v_2(p_T)$ is found to be less sensitive, but
the upper limit given by the eBC model, is moved closer to the RHIC results.

We also note that we expect the applicability region of hydrodynamical modeling
to extend higher in $p_T$ at the LHC than at RHIC, because pions from jet
fragmentation should start to dominate particle production over the
hydrodynamical spectra at higher $p_T$ at the LHC \cite{Eskola:2005ue}.
Therefore we expect that the minimum bias $v_2(p_T)$ will reach higher values at
the LHC even if the low-$p_T$ results are similar to those measured at RHIC.

\section*{Acknowledgments}
We thank the Academy of Finland, the project 115262, for financial support and 
P. Huovinen for discussions.

\end{document}